\documentclass[12pt]{article}

\usepackage{epsfig}
\usepackage{amsmath}
\usepackage{amssymb}
\usepackage{graphicx}
\usepackage{cite}
\usepackage{enumerate}
\usepackage[usenames,dvipsnames]{color}
\newcommand{\be}{\begin{equation}} 
\newcommand{\ee}{\end{equation}} 

\begin{document}
%\titlepage
\titlepage
\begin{flushright}
IFJPAN-IV-2012-4 \\
\end{flushright}
\vspace*{1in}
\begin{center}
{\Large \bf Resummation in nonlinear equation for high energy factorizable gluon density and its extension to include coherence.}\\
\vspace*{0.5cm}
Krzysztof Kutak \\
\vspace*{0.5cm}
 {\it Instytut Fizyki Jadrowej im H. Niewodniczanskiego,\\
Radzikowskiego 152, 31-342 Krakow, Poland\\}
\end{center}
\vspace*{1cm}
\centerline{(\today)}

\vskip1cm
\begin{abstract}
Motivated by forthcoming p-Pb experiments at Large Hadron Collider which require both knowledge of gluon densities accounting for 
saturation and for processes at a wide range of $p_t$ 
we study basic momentum space evolution equations of high energy QCD factorization. Solutions of those equations might be used to form a set 
of gluon densities to calculate observables in generalized high energy factorization. 
Moreover in order to provide a framework for predictions for exclusive final states in p-Pb scattering with high $p_t$ we rewrite the equation for the high energy factorizable
gluon density in a resummed form, similarly to what has been done in \cite{Kutak:2011fu} 
for the BK equation. The resummed equation is then extended to account for colour coherence. This introduces an external scale to the 
evolution of the gluon density, and therefore makes it applicable in studies of final states.
\end{abstract}
\section{Introduction}
One of the scientific plans for the LHC is to collide protons with lead nuclei.
This will offer an unprecedented possibility to test the partonic structure at high densities at relatively "cleaner" environment as compared 
to AA collisions \cite{Salgado:2011wc}.
The particularly interesting phenomenon which is believed to be understood with studies of observables produced in p-Pb is a phenomenon of gluon saturation
 \cite{Gribov:1984tu}. For recent overview we refer the reader to \cite{Kovchegov:2010pw,Gelis:2010nm}. 
Already some time ago it has been recognized that
saturation can be conveniently studied in the production of system of forward central di-jets \cite{Albacete:2010pg,Kharzeev:2004bw,Marquet:2007vb}, and recently that the signal of saturation might 
be expected in processes with high $p_t\simeq 20$ GeV or larger \cite{Kutak:2012rf}. 
With a system complicated enough like p-Pb, one may ask the question whether appropriate factorization of the process into hard parts, and parton 
densities can be establish taking all relevant degrees of freedom into account. To a good approximation i.e. neglecting the final state interactions of the forward jet one can use so called hybrid high energy factorization where one of parton densities parametrizes partons at large $x$ while the other one at low $x$ \cite{JalilianMarian:2005jf,Deak:2009xt}. However, it has been observed 
that generalized high energy factorization and appropriate hard matrix elements, and six gluon densities can be obtained providing more general framework.
The six gluon densities as has been shown in \cite{Dominguez:2011wm} can be expressed in terms of two basic gluon densities which can be 
obtained from the BK equation \cite{Balitsky:1995ub,Kovchegov:1999yj,Kovchegov:1999ua} originally formulated in the coordinate space.
Particularly useful for including higher order corrections, and numerical simulations of the scattering process is the momentum space 
representation/formulation of evolution equations. In this representation one can for example 
extend the BK equation to take into account sub leading pieces of the splitting function or bringing higher corrections in $\ln 1/x$ kinematical 
effects \cite{Kutak:2003bd,Peschanski:2006bm}. With momentum space formulation, one can, after performing 
resummation of the unresolved and virtual contributions in the kernel, extend the BK to include coherence which is a step towards studies of exclusive observables \cite{Kutak:2011fu}.
 
The resummation performed in \cite{Kutak:2011fu} transforms the full kernel (linear, and nonlinear part) of the equation to a form where so called Regge form factor multiplies full kernel. 
This result is highly non-trivial since, the equation was nonlinear, and the 
resummation technique was based on an integral transform method usually applicable to linear equations. The object for which the resummed equation was written was the so called Weizsacker-Williams gluon density (called in \cite{Kutak:2011fu} dipole amplitude in the momentum space). 
This is one of the two basic densities which are used to construct densities used in studies of di-jets as derived in \cite{Dominguez:2011wm}.

In the present article we investigate an equation of which the 
solution gives the so called high energy factorizable gluon density\cite{Catani:1990eg}. It turns out that although the nonlinear part of the 
equation for the unintegrated gluon density is much more complicated than the one studied in \cite{Kutak:2011fu}, one can perform still resummation
analogously to \cite{Kutak:2011fu}, the results
is the equation (\ref{eq:bkres}). The new form of the equation under consideration allows for the extraction of a 
more compact expression for the triple pomeron
vertex without splitting the vertex into real and virtual terms which will most probably be easier to handle in future numerical simulations. Although it is written in an exclusive 
form, obtained equation still does not allow for investigations of the exclusive processes since it does not include information on the hard process. The idea we follow is based on results
from \cite{Kutak:2011fu} where the BK equation was extended to include colour coherence by replacement of the Regge form factor with the non-Sudakov form factor, and the introduction
of angular ordering. The new result for the unintegrated gluon density with angular ordering is given by equation (\ref{eq:ccfmbk}). 
For other approach to extend the equation for the unintegrated gluon density to be able to deal with exclusive processes at high $p_t$ we refer the reader to \cite{Kutak:2003bd} and for latest applications of this framework to \cite{Kutak:2012rf}.

The paper is organized as follows. In section 2 we recall the evolution equations for the high energy factorizable unintegrated gluon density, and the BK equation, and we show explicitly how to transform one to the other. In section 3 we 
perform resummation of virtual and unresolved real contributions in the linear part of equation for the high energy factorizable gluon density. Furthermore, we 
extend the obtained equation to account for colour coherence. We conclude in section 4.

\section{Nonlinear equation for high energy factorizable unintegrated gluon density}
In this section we will review and collect some facts about the equation for high energy factorizable unintegrated gluon density. We also provide 
some new insight on implications of its solution.
The solution of the nonlinear evolution equation for the high energy factorizable unintegrated gluon density enters for example the formula for the
proton structure function, which at leading $\alpha_s\ln(1/x)$ reads:
\be
F_2(x,Q^2)=\frac{Q^2}{4\pi^2}\alpha_s\sum_q e_q^2\int d^2k\,{\cal F}(x,k^2)\left(S_L(k^2,Q^2,m_q^2)+S_T(k^2,Q^2,m_q^2)\right) 
\ee 
The impact factors $S_T$, and $S_L$ are obtained as matrix elements from the process $\gamma^*+g^*\rightarrow \overline q\,q$ with inclusion 
of appropriate phase space factors, the sum runs over quark flavours. In the massless case the full impact factor reads (see also 
\cite{Sergey:2008uh} for massive case):
\be
S_T(k^2,Q^2,0)+S_L(k^2,Q^2,0)=\int_0^1dz\int_0^1d\zeta\,\frac{1-2z(1-z)-2\zeta(1-\zeta)+12z(z-1)\zeta(1-\zeta)}{Q^2z(1-z)+k^2\zeta(1-\zeta)}
\ee
 
\begin{figure}[t] 
  \centering
  \includegraphics[width=0.3\textwidth]{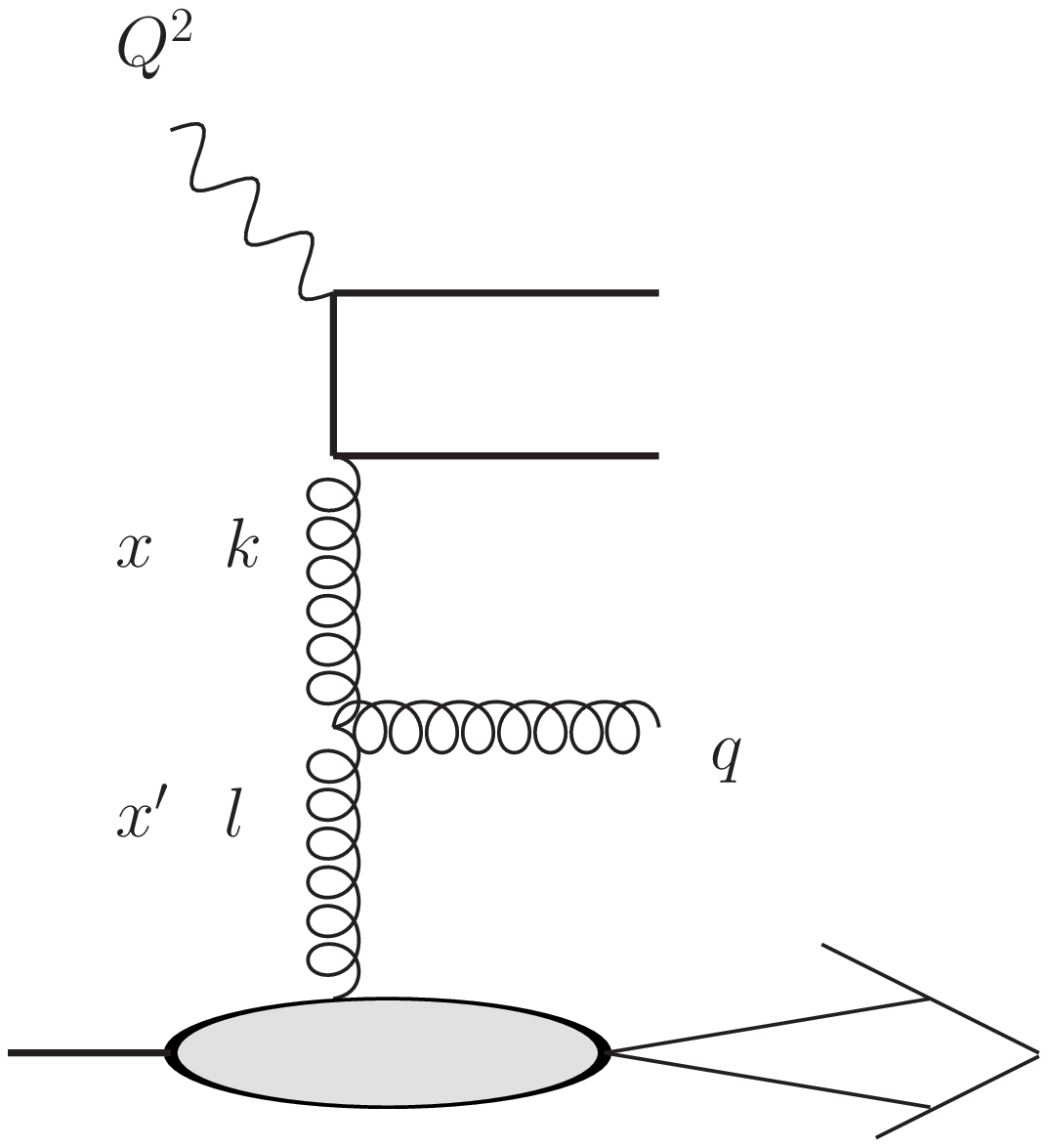}
  \includegraphics[width=0.3\textwidth]{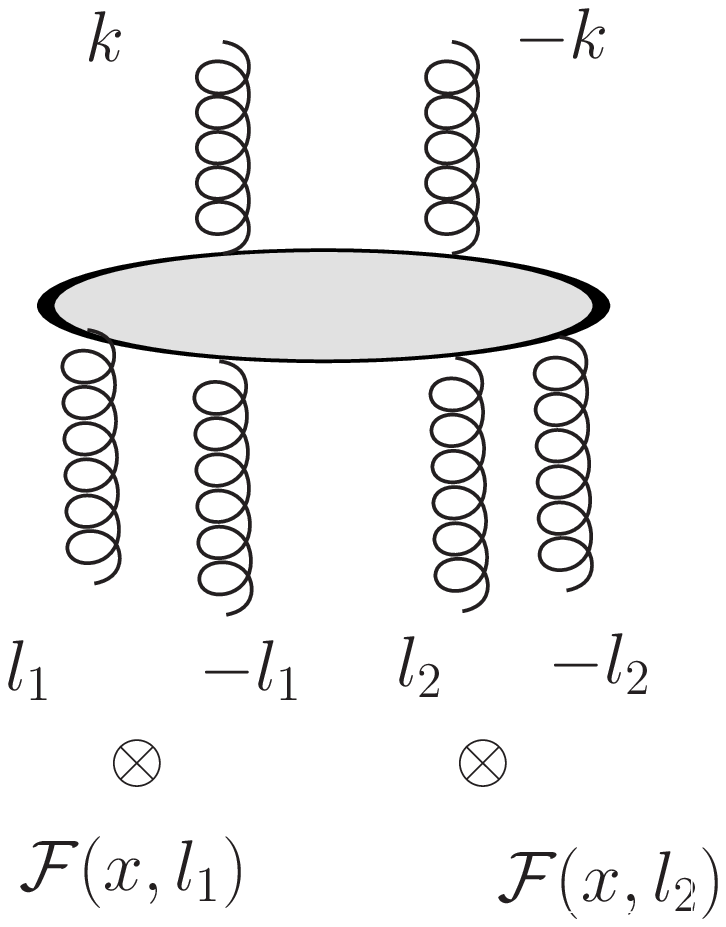}
 \caption{\em \small
  Left: Kinematical variables. 
  Right: Triple pomeron vertex convoluted with gluon density.
  }
  \label{fig:Kinematics} 
\end{figure}

For a full list of processes in which the gluon density $\cal{F}$ is applicable we refer the reader to \cite{Dominguez:2011wm}.\\
In the formulas above $Q^2$ is the four momentum square of the photon, $k^2$ is the virtuality of the gluon, which satisfies $k\equiv|{\bold k}|$, $l\equiv|{\bold l}|$ (where it will not lead to confusion we will not use for 2-d vectors bold font). Those momenta are lying in a plane transverse to the collision axis, 
(see Fig. (\ref{fig:Kinematics}) left).
At the leading order in $\alpha_s\ln 1/x$ the density ${\cal F}(x,k^2)$ obeys the following evolution equation \cite{Kutak:2003bd,Bartels:2007dm,Nikolaev:2006za}:
\be
\begin{split}
 {\cal F}(x,k^2)={\cal F}_0(x,k^2)+\overline\alpha_s\int_{x/x_0}^1\frac{dz}{z}\int_0^{\infty}\frac{dl^2}{l^2}
\bigg[\frac{l^2{\cal F}(x/z,l^2)- k^2{\cal F}(x/z,k^2)}{|k^2-l^2|}+ \frac{
k^2{\cal F}(x/z,k^2)}{\sqrt{(4l^4+k^4)}}\bigg]\\
-\frac{2\alpha_s^2\pi}{N_c R^2} \int_{x/x_0}^1\frac{dz}{z}\Bigg\{
\bigg[\int_{k^2}^{\infty}\frac{dl^2}{l^2}{\cal F}(x/z,l^2)\bigg]^2 
+\;{\cal F}(x/z,k^2)\int_{k^2}^{\infty}\frac{dl^2}{l^2}\ln\left(\frac{l^2}{k^2}\right){\cal F}(x/z,l^2)
\Bigg\}.
\label{eq:faneq1}
\end{split}
\ee
where $\overline\alpha_s=N_c\alpha_s/ \pi$ $z=x/x'$ (see Fig. (\ref{fig:Kinematics}) left). 
The nonlinear term in (\ref{eq:faneq1}) is a convolution of the angle-averaged triple pomeron vertex \cite{Bartels:2007dm} with a gluon density.  
 It reads:
\be
\begin{split}
{\cal V}(k,-k;l_1,-l_1,l_2,-l_2)=\frac{\pi\alpha_s^2}{N_c R^2}\Bigg[2\theta(l_1^2-k^2)\theta(l_2^2-k^2)
+k^2\ln\frac{l_1^2}{l_2^2}\delta(l_1^2-k^2)\theta(l_2^2-l_1^2)\\
+k^2\ln\frac{l_2^2}{l_1^2}\delta(l_2^2-k^2)\theta(l_1^2-l_2^2)\Bigg]
\end{split}
\ee
where the first term corresponds to the real contribution while the second and third to mixed virtual, and real terms. The arguments of the 
theta-functions indicate that the triple pomeron vertex orders the 
momenta in an anticollinear configuration.
Convolving it with the gluon density (see Fig. (\ref{fig:Kinematics}) right), propagators $1/l_1^2$, and $1/l_2^2$ (where remaining factors $1/l_1^2$, and $1/l_2^2$ were absorbed in 
the definition of gluon density), and changing variables we obtain the 
nonlinear term in equation (\ref{eq:faneq1}) (for more detailed derivation of (\ref{eq:faneq1}) we refer the reader to 
\cite{Bartels:2007dm}). 
The strength of the nonlinear term is controlled by the parameter $1/R^2$ which comes from the cut on a loop momentum circulating in the triple pomeron 
vertex when forward limit is 
taken. The inverse of a cut-off is interpreted as a hadron's radius \cite{Bartels:2007dm}. 
This equation can be transformed to coordinate space by rewriting the convolution formula for the $F_2$ structure function as a convolution of the dipole 
amplitude $N(r,b,x)$ and the photon wave function \cite{Kutak:2003bd}. Namely the relation between the dipole amplitude and high energy factorizable gluon density is \cite{Kharzeev:2003wz}:
\be
{\cal F}(x,k^2)=\frac{N_c}{\alpha_s (2\pi)^3}\int d^2b\int d^2re^{ik\cdot r}\nabla^2_{r}\,N(r,b,x)
\label{eq:transf}
\ee
For later applications we also write explicitly the relation between the Waizecker-Williams gluon density and the dipole amplitude. It reads
\be
\Phi(x,k^2)=\frac{1}{2\pi}\int d^2b \int\frac{d^2 r}{r^2}e^{i k \cdot r}N(r,b,x)
\label{eq:dipolampl}
\ee
From the relations above it follows that
\be
{\cal F}(x,k^2)=\frac{N_c}{4\alpha_s \pi^2}k^2\nabla^2_{k}\Phi(x,k^2)
\label{eq:laplaceop}
\ee
which in principle could be defined up to terms of type: 
\be
\Phi_C=C_1\ln k^2/\mu^2+C_2 
\label{eq:gauge}
\ee
but the relation of $\Phi(x,k^2)$ to the dipole amplitude (\ref{eq:dipolampl}) determines such terms to be zero. For more detailed discussion on this we relegate the Reader to the appendix.\\
We also note that relation (\ref{eq:laplaceop}) is in fact a Poisson equation allowing to some extend for an interpreting analogously 
to electrodynamics ${\cal F}$ as density while $\Phi$ as a potential. The inverse 
transformation can be performed since (\ref{eq:gauge}) does not contribute and the inverse is given by:
\be
\Phi(x,k^2)=\frac{\alpha_s\pi^2}{N_c}\int_{k^2}^{\infty}\frac{dl^2}{l^2}\ln\frac{l^2}{k^2}{\cal F}(x,l^2).
\ee
The function $\Phi(x,k^2)$ is linked to a function $\Phi_b(x,k^2)$ which obeys the BK equation (in momentum space):
\be
\begin{split}
\Phi_b(x,k^2)= \Phi_{0\,b}(x,k^2)+\overline\alpha_s\int_{x/x_0}^1\frac{dz}{z}
\int_0^{\infty}\frac{dl^2}{l^2}
\bigg[\frac{l^2\Phi_b(x/z,l^2)- k^2\Phi_b(x/z,k^2)}{|k^2-l^2|}+ \frac{
k^2\Phi_b(x/z,k^2)}{\sqrt{(4l^4+k^4)}}\bigg]\\
-\overline\alpha_s\int_{x/x_0}^1\frac{dz}{z}\Phi_b^2(x/z,k^2).
\label{eq:wwglue0}
\end{split}
\ee
In the equation above, impact parameter is not a dynamical quantity. Furthermore (\ref{eq:wwglue0})
does not describe physics at large distances correctly since it does 
not take into account confinement effects \cite{Kovner:2002yt}. Taking above into account in order to obtain the integrated over $b$ gluon density one has to assume some
impact parameter dependence since one needs to perform the integration over it to calculate gluon density. For example assuming cylinder like target one may use the following anzatz for factorization 
\cite{Kutak:2003bd}

\be
\Phi_b(x,k^2)=\Phi(x,k^2)S(b).
\ee
with normalization conditions
\be
\int d^2{\bf b}\,S(b)=1,\,\,\,\int d^2b\,S^2(b)=\frac{1}{\pi R^2}
\ee
After integration over the impact parameter $b$ we obtain:
\be
\begin{split}
\Phi(x,k^2)= \Phi_0(x,k^2)+\overline\alpha_s\int_{x/x_0}^1\frac{dz}{z}
\int_0^{\infty}\frac{dl^2}{l^2}
\bigg[\frac{l^2\Phi(x/z,l^2)- k^2\Phi(x/z,k^2)}{|k^2-l^2|}+ \frac{
k^2\Phi(x/z,k^2)}{\sqrt{(4l^4+k^4)}}\bigg]\\
-\frac{\overline\alpha_s}{\pi R^2}\int_{x/x_0}^1\frac{dz}{z}\Phi^2(x/z,k^2).
\label{eq:wwglue}
\end{split}
\ee
where upon the integration over the impact parameter the inverse radius of the proton $R$ multiplies the nonlinear term in (\ref{eq:wwglue}).\\
The transformation between (\ref{eq:faneq1}) and (\ref{eq:wwglue}) is known, however it can be performed in a more straight forward way than in \cite{Kutak:2007vf} and is presented below. 
The linear part of the kernel can be written in the form of a differential operator \cite{Kovchegov:1999ua}. We have:
\be
\Phi(x,k^2)= \Phi_0(x,k^2)+\overline\alpha_s\int_{x/x_0}^1\frac{dz}{z}\chi\left(-\frac{\partial}{\partial_{\log k^2}}\right)\Phi(x/z,k^2)-
\frac{\overline\alpha_s}{\pi R^2}\int_{x/x_0}^1\frac{dz}{z}\Phi^2(x/z,k^2)
\label{eq:BKdip}
\ee
where $\chi(\gamma)=2\psi(1)-\psi(\gamma)-\psi(1-\gamma)$ is the characteristic function of the BFKL.
We represent the BFKL kernel of the BK equation as a power series around $\gamma_c=0.373$:
\be
\chi(-\partial_{\ln k^2})\Phi(x,k^2)=\left[\chi(\gamma_c)+(-\partial_{\ln k^2}-\gamma_c)\chi'(\gamma_c)+\frac{1}{2!}(-\partial_{\ln k^2}-\gamma_c)^2\chi''(\gamma_c)+...\right]\Phi(x,k^2)
\label{eq:BK}
\ee 
%%%%%%%%%%%%%%%%%%%%%%%%%%%%%%%%%%%%%%%%%%%%%%%%%%%%%%%%%%%%%%%%%%%
\begin{figure}[t!]
  \begin{picture}(30,30)
    \put(30, -80){
      \includegraphics{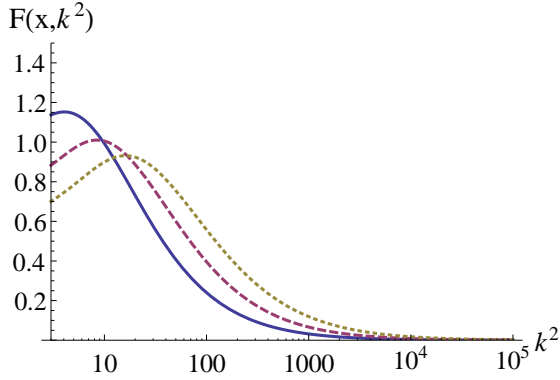}
    }

      \end{picture}
\vspace{3cm}
\caption{\em \small High energy factorizable gluon density. The maxima signalize emergence of the saturation scale $Q_s(x)$. Plot is made for $x=10^{-5}$ (continuous line), $x=10^{-6}$ (dashed line), $x=10^{-7}$ (dotted line)}
%\vspace{7cm}
\label{fig:plotvel}
\end{figure}
Inserting this expansion into (\ref{eq:BKdip}), and noticing that the BFKL kernel is invariant with respect to 
$k^2\partial_{k^2}k^2\partial_{k^2}\equiv\partial_{\ln k^2}^2$ we obtain the linear part of 
the equation for the high energy factorizable unintegrated gluon density in the same form as for the WW gluon density. In order to invert the nonlinear part we use 
(\ref{eq:transf}). Folding the linear kernel again we obtain:
\be
\begin{split}
{\cal F}(x,k^2)= {\cal F}_0(x,k^2)+\overline\alpha_s\int_{x/x_0}^1\frac{dz}{z}
\int_0^{\infty}\frac{dl^2}{l^2}
\bigg[\frac{l^2{\cal F}(x/z,l^2)- k^2{\cal F}(x/z,k^2)}{|k^2-l^2|}+ \frac{
k^2{\cal F}(x/z,k^2)}{\sqrt{(4l^4+k^4)}}\bigg]\\
-\frac{\pi\alpha_s^2k^2}{4N_cR^2} \nabla_k^2\int_{x/x_0}^1\frac{dz}{z}\left[\int_{k^2}^{\infty}\frac{dl^2}{l^2}\ln\frac{l^2}{k^2}{\cal F}(x/z,l^2)\right]^2
\label{eq:faneq}
\end{split}
\ee
where the nonlinear part is written in a new compact form.
The equation above has quite interesting properties. In particular the main contribution to the nonlinear part comes from the anticollinear part 
of the triple pomeron vertex (recall the theta functions in the triple
 pomeron vertex). This has been already observed in \cite{Bartels:2007dm} but the consequences of this finding were not fully clear. 
Since the nonlinear part enters the equation with a minus sign, the anticollinear pole is subtracted from the linear kernel during evolution. The subtraction is stronger when the $x$ is smaller where the importance of the nonlinear part of the equation is larger. 
The presence of the collinear and anticollinear contributions to the linear kernel of (\ref{eq:faneq}) give rise to unordered emissions. The subtractions 
of the anticollinear pole via triple pomeron vertex result in the collinear ordering of the gluonic chain both in $x$, and in transversal 
momentum's most probable value which is given by emergent saturation scale $Q_s$. This can be seen by inspecting the shape of the gluon density as 
function of $k$. The plot (\ref{fig:plotvel}) shows that the lower values of $x$ the saturation scale is harder and the distribution shifts towards higher values
of the transversal momentum of a gluon with its maximal value at the saturation scale $Q_s$. The property of providing most probable 
value is a feature that distinguishes the unintegrated gluon density obtained 
from (\ref{eq:BK}) from unintegrated distributions obtained from linear evolution equations or from (\ref{eq:wwglue}).  The most probable momentum in the latter case 
is given by its boundary value, since the distribution diverges at small $k$, and some cut has to be applied.
%%%%%%%%%%%%%%%%%%%%%%%%%%%%%%%%%%%%%%%%%%%%%%%%%%%%%%%%%%%%%%%%%%%%%%%%%%%%%%%%%%%%%
\section{Resummed form of the equation for high energy factorizable gluon density}
In order to find the resummed form of (\ref{eq:faneq}) we follow similar steps as in \cite{Kutak:2011fu}. 
In principle as described above, (\ref{eq:wwglue}) and (\ref{eq:faneq}) are related via the Laplace operator so in principle after solving (\ref{eq:wwglue}) one obtains (\ref{eq:faneq}).
However, for practical future Monte Carlo simulation of $p\,+\,Pb\to di\!-\!jets +X$ where both of these densities enter it is more convenient to have separate equations.   
Another reason for having both equations is that, the dynamics of gluon evolution is more transparent in (\ref{eq:faneq}) since one explicitly sees the role of the triple pomeron vertex in bringing the unitarity corrections via removing the anticollinear pole. Also the limiting cases like the double logarithmic limit is easily seen in the later.   
In the first step the angle dependence in the linear part (\ref{eq:faneq}) has to be reintroduced, and virtual corrections 
have to be collected in a one expression. After this procedure one obtains 
an exclusive representation of the linear part, while the nonlinear we leave unchanged. Thus we have
\be
\begin{split}
\label{eq:bkversion1}
{\cal F}(x,k^2)= {\cal F}_{0}(x,k^2)+\overline\alpha_s\int_{x/x_0}^1\frac{dz}{z}\Bigg\{\int\frac{d^2{\bf q}}{\pi q^2}\big[{\cal F}(x/z,|{\bold k}+{\bold q}|^2)-\theta(k^2-q^2){\cal F}(x/z,k)\big]\\
-\frac{\pi\alpha_s^2k^2}{4N_cR^2} \nabla_k^2\left[\int_{k^2}^{\infty}\frac{dl^2}{l^2}\ln\frac{l^2}{k^2}{\cal F}(x/z,l^2)\right]^2\Bigg\}
\end{split}
\ee

In the next step the linear part has to be decomposed into so called resolved, unresolved ones, and virtual emissions. The splitting is 
dictated by a singularity appearing in the denominator in the linear part of 
(\ref{eq:bkversion1}) when $q\rightarrow 0$.
The parameter which controls this splitting $\mu$ satisfies the following inequalities: $q\!>\!\mu$ for the resolved part , and $q\!<\!\mu$ 
for the unresolved part.
Thus we have
\begin{align}
\label{eq:bkversion2}
{\cal F}(x,k^2)&={\cal F}_0(x,k^2)\\\nonumber
&+\overline\alpha_s\int_{x/x_0}^1\frac{dz}{z}\int\frac{d^2{\bf q}}{\pi q^2}{\cal F}(x/z,|{\bold k}+{\bold q}|^2)\theta(q^2-\mu^2)\\\nonumber
&+\overline\alpha_s\int_{x/x_0}^1\frac{dz}{z}\int\frac{d^2{\bf q}}{\pi q^2}\big[{\cal F}(x/z,|{\bold k}+{\bold q}|^2)\theta(\mu^2- q^2)-
\theta(k^2-q^2){\cal F}(x/z,k)\big]\\\nonumber
&-\frac{\pi\alpha_s^2k^2}{4N_c\,R^2} \nabla_k^2\left[\int_{k^2}^{\infty}\frac{dl^2}{l^2}\ln\frac{l^2}{k^2}{\cal F}(x/z,l^2)\right]^2\,.
\nonumber
\end{align}
To perform resummation of virtual, and unresolved real emissions we are going to use the Mellin transform, defined as
\be
\overline{\cal F}(\omega,k^2)=\int_0^{x_0} dx x^{\omega-1} {\cal F}(x,k^2)
\ee
while the inverse transform reads
\be
{\cal F}(x,k^2)=\int_{c-i\infty}^{c+i\infty} d\omega\, x^{-\omega} \overline{\cal F}(\omega,k^2)\,.
\label{eq:invmellin}
\ee
Performing the Mellin transform, and using in the unresolved part $|{\bold k}+{\bold q}|^2\approx{\bold k}^2$ since $q^2<\mu^2$ we obtain
\begin{align}
\label{eq:eqtransform}
\overline {\cal F}(\omega,k^2)&=\overline {\cal F}_{0}(\omega,k^2)\\\nonumber
&+\frac{\overline\alpha_s}{\omega}\int\frac{d^2{\bold q}}{ q^2}[\overline {\cal F}(\omega,|{\bf k} +{\bf q}|^2)\theta(q^2-\mu^2)]
+\frac{\overline\alpha_s}{\omega}\int\frac{d^2{\bold q}}{ q^2}{\overline {\cal F}}(\omega,k^2)[\theta(\mu^2-q^2)-\theta(k^2-q^2)]\\\nonumber
&-\frac{\overline{\alpha}_s^2}{R^2}\frac{1}{\omega}\int_0^1 dy y^{\omega-1}k^2\nabla_k^2\left[\int_{k^2}^{\infty}\frac{dl^2}{l^2}
\ln\frac{l^2}{k^2}{\cal F}(x/z,l^2)\right]^2
\end{align}
in the nonlinear term we changed the variables $x/z\rightarrow y$, and we integrated over $x$ giving $1/\omega$ in front of 
the nonlinear part.
We obtain
\begin{align}
\overline {\cal F}(\omega,k^2)&=\overline {\cal F}_{0}(\omega,k^2)\\\nonumber
&+\frac{\overline\alpha_s}{\omega}\int\frac{d^2{\bf q}}{\pi q^2}\overline {\cal F}(\omega,|{\bf k} +{\bf q}|^2)\theta(q^2-\mu^2)-
\frac{\overline\alpha_s}{\omega}\overline {\cal F}(\omega,k^2)\ln\frac{k^2}{\mu^2}\\\nonumber
&-\frac{\pi\alpha_s^2}{4N_c\,R^2\omega}\int_0^1 dy y^{\omega-1}k^2 \nabla_k^2\left[\int_{k^2}^{\infty}\frac{dl^2}{l^2}\ln\frac{l^2}{k^2}{\cal F}(\omega,l^2)\right]^2\,.
\nonumber
\end{align}
simplifying we obtain
\be
\begin{split}
\overline {\cal F}(\omega,k^2)=\hat {\cal F}_{0}(\omega,k^2)+ 
\frac{\overline\alpha_s}{\overline\omega+\omega}\int\frac{d^2{\bf q}}{\pi q^2}[\overline {\cal F}(\omega,|{\bf k}+{\bf q}|^2)]
\theta(q^2-\mu^2)\\
-\frac{1}{\overline\omega+\omega}\frac{\pi\alpha_s^2}{4N_cR^2}\int_0^1 dy y^{\omega-1}k^2 
\nabla_k^2\left[\int_{k^2}^{\infty}\frac{dl^2}{l^2}\ln\frac{l^2}{k^2}{\cal F}(\omega,l^2)\right]^2
\end{split}
\ee

where
\be
\hat {\cal F}_{0}(\omega,k^2)=\frac{\omega\,\overline{\cal F}_{0}(\omega,k^2)}{\omega+\overline\omega}\,,~~~
\overline\omega=\overline\alpha_s\ln\frac{k^2}{\mu^2}.
\ee

%The inverse transform is defined as:
%\be
%{\cal F}(x,k^2)=\int_{c-i\infty}^{c+i\infty} d\omega x^{-\omega} \overline{\cal F}(\omega,k^2)
%\ee
Inverting the transform 
we arrive at:
\begin{align}
\label{eq:bkres}
{\cal F}(x,k^2)=\tilde {\cal F}_0(x,k^2)+&\overline\alpha_s\int_{x/x_0}^1\frac{d\,z}{z}\Delta_R(z,k,\mu)\Bigg\{\int\frac{d^2{\bf q}}{\pi q^2}\theta(q^2-\mu^2){\cal F}(\frac{x}{z},|{\bf k}+{\bf q}|^2)\\
&-\frac{\pi\alpha_s^2}{4N_cR^2}k^2\nabla_{k}^2\left[\int_{k^2}^{\infty}\frac{dl^2}{l^2}\ln\frac{l^2}{k^2}{\cal F}(x,l^2)\right]^2\Bigg\}\nonumber
\end{align}
where $\Delta_R(z,k,\mu)\equiv\exp\left(-\overline\alpha_s\ln\frac{1}{z}\ln\frac{k^2}{\mu^2}\right)$ is the Regge form factor and we defined
\be
\tilde{\cal F}_0(x,k^2)\equiv\frac{1}{2\pi i}\int_{c-i\infty}^{c+i\infty}d\omega x^{-\omega}\hat{\cal F}^0(\omega,k^2)
\ee
From equation (\ref{eq:bkres}) we may extract the new form of the triple pomeron vertex. 
\be
{\cal V}_{\mathrm{resummed}}=\frac{\pi\alpha_s^2}{4\,N_c\,R^2}k^2\Delta_R(z,k,\mu)\nabla_{k}^2\ln\frac{l_1^2}{k^2}\theta(l_1^2-k^2)\ln\frac{l_2^2}{k^2}\theta(l_2^2-k^2)
\ee
In the form above it is understood that the Laplacian has to be taken after convolution with the parton density is performed.\\
The last equation may be generalized to account for colour coherence,
replacing the Regge form factor by the non-Sudakov form factor, and by introducing angular ordering in an angle of emitted gluons with respect to the beam direction. 
The extended equation takes the form:
\begin{align}
\label{eq:ccfmbk}
{\cal F}(x,k^2,p)=\tilde {\cal F}_0(x,k^2,p)\\\nonumber
&+\overline\alpha_s\int\frac{d^2{\bf q}}{\pi q^2}\int_{x/x_0}^1\frac{d\,z}{z}\theta(p-q\,z)\Delta_{ns}(z,k,q)\Bigg\{{\cal F}(\frac{x}{z},|{\bf k}+{\bf q}|^2,q)\\\nonumber
&-\frac{\pi\alpha_s^2}{4N_cR^2}q^2\delta(q^2-k^2)\nabla_{q}^2\left[\int_{q^2}^{\infty}\frac{dl^2}{l^2}\ln\frac{l^2}{k^2}{\cal F}(x/z,l^2,l)\right]^2\Bigg\}
\end{align}
The momentum vector associated with $i$-th emitted gluon is
\be
q_i=\alpha_i\,p_P+\beta_i\,p_e+q_{t\,i}
\ee
and thus we have rapidity and angle of emitted gluon with respect to incoming parent proton (beam direction)
\be
\eta_i=\frac{1}{2}\ln(\xi_i)\equiv\frac{1}{2}\ln\left(\frac{\beta_i}{\alpha_i}\right)=\ln\left(\frac{|{\bf
q}_{i}|}{\sqrt{s}\,\alpha_i}\right),\,\,\,\,\,\tan\frac{\theta_i}{2}=\frac{|{\bf q}_i|}{\sqrt s\,\alpha_i}\,.
\ee
The variable $p$ in (\ref{eq:ccfmbk}) is defined via $\bar{\xi} = p^2/(x^2s)$ where $\frac{1}{2}\ln(\bar{\xi})$ is a maximum 
rapidity which is determined by the kinematics of the hard scattering,
$\sqrt{s}$ is the total energy of the collision. 
The particularly compact form of (\ref{eq:ccfmbk})
where the $\Delta_{ns}$ acts both on the linear, and the nonlinear piece might be a subject to further extensions. For example one can use a non-Sudakov form factor which takes into account so 
called kinematical effects which bring important pieces of higher order corrections to those of $\ln(1/x)$ type. Another possibility is to account for sub leading pieces of the splitting function therefore proposing the extension of the CCFM equation \cite{Ciafaloni:1987ur,Catani:1989sg,Catani:1989yc,Marchesini:1994wr} for unintegrated gluon density with nonlinearities.

The unintegrated gluon distributions obtained from equation (\ref{eq:bkres}) corresponds to the coordinate space dipole in the 
fundamental representation. For LHC applications one 
also needs it in the adjoin representation. The corresponding unintegrated gluon density in the adjoin representation can be 
obtained by calculating the corresponding dipole cross section, and transforming it back to momentum space. To be precise one applies
\be
\sigma(x,k^2)=\frac{8\pi^2}{N_c}\int\frac{dk}{k}(1-J_0(x,k^2)){\cal F}(x,k^2)
\label{eq:dipolexsection}
\ee   

\noindent together with
\be
N_G(x,r,b)=2N(x,r,b)-N^2(x,r,b),\,\,\,\sigma(x,r)=2\int d^2bN(x,r,b)
\ee

\noindent to obtain the gluon density in the colour adjoin representation:
\be
{\cal F}_G(x,k^2)=\frac{C_F}{\alpha_s(2\pi)^3}\int d^2b\int d^2r e^{i r\cdot k}\nabla_r^2 N_G(x,r,b)
\ee
After proposing equation (\ref{eq:ccfmbk}) one can generalize the formulas above to the situation where the dipole density depends on the 
hardness. In this case formula (\ref{eq:dipolexsection}) changes to:
\be
\sigma(x,r,p)=\frac{8\pi^2}{N_c}\int\frac{dk}{k}(1-J_0(x,k^2)){\cal F}(x,k^2,p)
\ee

\noindent In a similar way equations below (\ref{eq:dipolexsection}) can be generalized to account for hardness. 
We have
\be
{\cal F}_G(x,k^2,p)=\frac{C_F}{\alpha_s(2\pi)^3}\int d^2b\int d^2r e^{i r\cdot k}\nabla_r^2 N_G(x,r,p,b)
\ee
It will be interesting once the equation (\ref{eq:ccfmbk}) is solved to check the behaviour of the dipole amplitude at large $r$, and in particular whether it flattens at constant value for different $p$.

\section{Conclusions, and outlook}

In this paper we studied evolution equations for the gluon density, the solutions of which form the basis for gluon densities entering the 
generalized high energy factorization framework. 
We have shown explicitly how the equations for the Weizsacker-Williams gluon density, and the high energy factorizable gluon density are related to 
each other. In the next step we performed resummation 
of the unresolved real emissions in the evolution equation for high energy factorizable gluon density. The interesting aspect of our result is that the triple pomeron
 vertex itself does not need to be resummed since it is no singular, but the resummation of the linear part affects it in a simple 
 multiplicative way. In the next step we proposed an 
 equation which generalizes the resumed equation to account for colour coherence. The last equation is interesting because of possible future applications in studies of high $p_t$ exclusive final states with a focus on saturation. In the future it will be interesting to see via numerical calculations how the resummations performed affects both of investigated gluon densities i.e. how solutions of equations before resummation are related to solutions after resummation. On the level of linear equations it has been shown that the resummed equation gives for small enough $\mu$ identical answer to the solution before resummation \cite{Schmidt:1996fg} and that the $\mu$ dependence is an effect of higher twist \cite{Marchesini:1994wr}. However, in the case of nonlinear equations the nonlinear part introduces effects of higher twists and therefore there might be some nontrivial interplay, for small enough saturation scale, of the saturation scale $Q_s$ and the cut-off scale $\mu$.  In will be also interesting to study to what extend the resummation affects the relations via Laplace operator between gluon densities i.e. what  is the relation of direct solution of (\ref{eq:bkres}) to action of laplacian on solution of resummed version of (\ref{eq:wwglue}).
\section*{Acknowledgments}
I would like to thank Krzysztof Golec-Biernat for useful comments on the manuscript and Andreas van Hameren for proofreading. Discussions with Stanislaw Jadach, Wieslaw Placzek, Sebastian Sapeta, Maciej Skrzypek are kindly acknowledged.
This research has been supported by NCBiR grant LIDER/02/35/L-2/10/NCBiR/2011
\section{Appendix}
In this appendix we will perform the inversion of the Laplace operator in (\ref{eq:laplaceop}) and provide an example of action of it's inverse on the high energy factorizable gluon density within the GBW saturation model. We will also show how the behavior of the dipole amplitude sets constraint on the WW gluon density.
The equation (\ref{eq:laplaceop}) reads:
\be
{\cal F}(x,k^2)=\frac{N_c}{4\alpha_s \pi^2}k^2\nabla^2_{k}\Phi(x,k^2)
\label{eq:laplaceop12}
\ee
The physical boundary condition that the WW gluon density obeys in the highly perturbative domain i.e. $k^2\rightarrow\infty$ is that $\Phi(x,k^2)|_{k^2=\infty}=0$ and $k^2\partial_{k^2}\Phi(x,k^2)|_{k^2=\infty}=0$. It might be linked to the behaviour of the dipole cross section for small dipole size
\be
\Phi(x,k^2)=\int d^2b \int_0^{\infty}\frac{dr}{r}J_0(k\,r))N(x,r,b)=
\int d^2b\int_0^{1/k^2}\frac{dr^2}{r^2}r^2=S_\perp\frac{1}{k^2}
\label{eq:appendix1}
\ee
where we used $N(x,r,b)\approx\theta(b_0-b)\,r^2$ and $J(0,k\,r)\approx 1$ for small $r$ \cite{Kharzeev:2003wz,KovchegovLevin}. The theta function indicates that we model a hadron as a cylinder, the integral over the impact parameter $b$ gives targets area $S_\perp$ while the $r^2$ behaviour of the dipole amplitude comes from colour transparency.
The behaviour of the dipole amplitude at small $r$ is sufficient to set the terms $C_1$ and $C_2$ to zero in (\ref{eq:gauge}) since  
they would modify the gluon density globally i.e. the term $\ln k^2/\mu^2$ would lead to a divergent behaviour of the WW gluon density at large $k$ while the $C_2$ would introduce a constant contribution over the whole region which is also not consistent with $1/k^2$ behaviour of the gluon density. Setting this terms to zero is therefore justified and allows for inversion of the Laplace operator \cite{Fuller} with what we proceed below.
The equation (\ref{eq:laplaceop12}) can be rewritten as 
\be
{\cal F}(x,u)=\frac{N_c}{4\alpha_s \pi^2}\partial_u^2\Phi(x,u)
\label{eq:laplaceop2}
\ee
where $u=\ln k^2/k_0^2$ we set $k_0^2=1$ and denote $A\equiv\frac{N_c}{4\alpha_s \pi^2}$.
Integrating once and using the boundary condition $\partial_u\Phi(x,u=\infty)=0$ we obtain:
\be
\partial_u\Phi(x,u)=-\frac{1}{A}\int_u^{\infty}du^{\prime}{\cal F}(x,u^{\prime})
\ee
Integrating one more time and using the  boundary condition for the function $\Phi(x,u=\infty)=0$ we obtain:
\be
\Phi(x,u)=\frac{1}{A}\int_{u}^\infty du^{\prime}{\cal F}(x,u^{\prime})(u^{\prime}-u)
\ee
Using the original variables we have
\be
\Phi(x,k^2)=\frac{\alpha_s\pi^2}{N_c}\int_{k^2}^{\infty}\frac{dl^2}{l^2}\ln\frac{l^2}{k^2}{\cal F}(x,l^2).
\ee
Now we proceed with the example of the application of the above formula to high energy factorizable gluon density ${\cal F}(x,k^2)$. As an example we take the GBW gluon density \cite{GolecBiernat:1999qd}. In this model the gluon density falls too fast at large $k^2$ as compared to perturabative QCD pattern but nevertheless it models quite well bulk properties of physics of saturation. It has also correct asymptotics and is frequently used as initial distributions for evolution equations \cite{Sergey:2008uh,Enberg:2005cb}. 
Applying the formula (\ref{eq:laplaceop2}) to ${\cal F}(x,k^2)=\frac{N_cS_\perp}{2\pi^2\alpha_s}k^2/Q_s^2(x)e^{-k^2/Q_s^2(x)}$ we obtain
\be
\Phi(x,k^2)=\frac{S_\perp}{2}\Gamma(0,k^2/Q_s^2)
\label{eq:appendix2}
\ee
which satisfies the asymptotic behaviour $\Phi(x,k^2)\simeq 0$ for $k^2\rightarrow\infty$.
The dipole cross section in the GBW model follows from inserting (\ref{eq:appendix2}) to inverse of (\ref{eq:dipolampl}) and reads
\be
\sigma(x,r^2)=S_\perp r^2\int \frac{d^2k}{2\pi}e^{-ik\cdot r}\Phi(x,k^2)=2S_\perp\left(1-e^{\frac{r^2\,Q_s^2(x)}{4}}\right)
\ee
in agreement with \cite{GolecBiernat:1998js}.

Alternatively to what we show above in order to determine the boundary conditions at large $k^2$ of the WW gluon density we can start with the lowest order expression in $\alpha_s\ln (1/x)$ for high energy factorizable gluon density i.e. 
${\cal F}(x,k^2)=const\frac{1}{k^2}$.
Using  (\ref{eq:laplaceop2}) we have
\be
\Phi(x,k^2)=\frac{\alpha_s\pi^2}{N_c}\int_{k^2}^{\infty}\frac{dl^2}{l^2}\ln\frac{l^2}{k^2}{\cal F}(x,k^2)=const\frac{\alpha_s\pi^2}{N_c}\frac{1}{k^2}
\ee
what is consistent with (\ref{eq:appendix1}) and (\ref{eq:appendix2}) and also justifies setting terms $C_1$ and $C_2$ to zero.

\end{document}